\documentclass[twocolumn,showpacs,aps,prl,superscriptaddress]{revtex4}

\usepackage{graphicx}
\usepackage{dcolumn}
\usepackage{amsmath}
\usepackage{epsfig}

\input pubboard/babarsym

\def\pipi     {\ensuremath{\pi\pi}}

\def\hh     {\ensuremath{h^+h^{\prime -}}}

\def\fpm {\ensuremath{f_{\pm}(\deltat)}}
\def\ilam {\ensuremath{{\cal I}m\lambda}}
\def\alam {\ensuremath{\left|\lambda\right|}}

\def\spipi {\ensuremath{S_{\pi\pi}}}
\def\cpipi {\ensuremath{C_{\pi\pi}}}
\def\de {\ensuremath{\Delta E}}

\def\diffD {\ensuremath{\Delta D}}

\def\Btag {\ensuremath{B_{\rm tag}}}
\def\Bhh {\ensuremath{B_{hh}}}
\def\Bflav {\ensuremath{B_{\rm flav}}}

\def\ttag {\ensuremath{t_{\rm tag}}}

\newcommand{\BABARPubYear}    {01}
\newcommand{\BABARPubNumber}  {21}

\newcommand{\SLACPubNumber} {9012}

\def\figurebox#1#2#3{%
    \def\arg{#3}%
    \ifx\arg\empty
    {\hfill\vbox{\hsize#2\hrule\hbox to #2{\vrule\hfill\vbox to #1{\hsize#2\vfill}\vrule}\hrule}\hfill}%
    \else
    {\hfill\epsfbox{#3}\hfill}%
    \fi}

\begin{document}

\preprint{\babar-PUB-\BABARPubYear/\BABARPubNumber} 
\preprint{SLAC-PUB-\SLACPubNumber} 

\begin{flushleft}
\babar-PUB-\BABARPubYear/\BABARPubNumber\\
SLAC-PUB-\SLACPubNumber\\[10mm]
\end{flushleft}

\title{
{\large \bf
Study of $\CP$-violating asymmetries in {\boldmath $\Bz\to\pip\pim,\,\Kp\pim$} decays} 
}

%
\author{B.~Aubert}
\author{D.~Boutigny}
\author{J.-M.~Gaillard}
\author{A.~Hicheur}
\author{Y.~Karyotakis}
\author{J.~P.~Lees}
\author{P.~Robbe}
\author{V.~Tisserand}
\affiliation{Laboratoire de Physique des Particules, F-74941 Annecy-le-Vieux, France }
\author{A.~Palano}
\author{A.~Pompili}
\affiliation{Universit\`a di Bari, Dipartimento di Fisica and INFN, I-70126 Bari, Italy }
% ...
\collaboration{The \babar\ Collaboration}
\noaffiliation

\date{October 25, 2001}

\begin{abstract}
We present a measurement of the time-dependent $\CP$-violating
asymmetries in neutral $B$ decays to the $\pip\pim$ \CP\ eigenstate, and 
an updated measurement of the charge asymmetry in $\Bz\to \Kp\pim$ decays.  
In a sample of $33$ million $\FourS\to\BB$ decays collected with the 
\babar\ detector at the SLAC PEP-II asymmetric $B$ Factory, we
find $65^{+12}_{-11}$ $\pip\pim$ and $217\pm 18$ $\Kp\pim$ candidates
and measure the asymmetry parameters $\spipi = 0.03^{+0.53}_{-0.56}\pm 0.11$, 
$\cpipi = -0.25^{+0.45}_{-0.47}\pm 0.14$, and ${\cal A}_{K\pi} = -0.07 \pm 0.08 \pm 0.02$, 
where the first error is statistical and the second is systematic.
\end{abstract}


\maketitle

In the Standard Model, all $\CP$-violating effects arise from a single
complex phase in the three-generation Cabibbo-Kobayashi-Maskawa (CKM)
quark-mixing matrix~\cite{CKM}.  One of the central questions in particle 
physics is whether this
mechanism is sufficient to explain the pattern of \CP violation observed 
in nature.  Recent measurements of the parameter $\stwob$ by the
\babar\/~\cite{BabarSin2betaObs} and BELLE~\cite{BelleSin2betaObs} Collaborations 
establish that \CP\ symmetry is violated in the neutral $B$-meson system.  
In addition to measuring $\stwob$ more precisely, one of the primary goals of the 
$B$-Factory experiments in the future will be to measure the remaining angles 
($\alpha$ and $\gamma$) and sides of the Unitarity Triangle in order to further 
test whether the Standard Model description of \CP\ violation is correct.

The study of $B$ decays to charmless hadronic two-body final states
will play an increasingly important role in our understanding of \CP violation.  
In the Standard Model, the time-dependent $\CP$-violating asymmetry in
the reaction $\Bz\to\pip\pim$ is related to the angle $\alpha$.  
In addition, observation of a significant rate asymmetry between 
$\Bz\to\Kp\pim$ and $\Bzb\to\Km\pip$ decays would be evidence for direct 
\CP violation, and ratios of branching fractions for various $\pi\pi$ 
and $K\pi$ decay modes are sensitive to the angle $\gamma$.  Finally,
branching fraction measurements provide critical tests of theoretical 
models that are needed to extract reliable information on \CP\ violation
from the experimental observables.

The \babar\ Collaboration recently reported measurements of branching
fractions and charge asymmetries for several charmless two-body $B$ decays 
using a data set of $23$ million $\BB$ pairs~\cite{twobodyPRL}.  
In this paper, using a data sample of approximately $33$ million 
$\BB$ pairs, we report a measurement of time-dependent 
$\CP$-violating asymmetries in neutral $B$ decays to the $\pip\pim$ 
\CP\ eigenstate and an updated measurement of the charge asymmetry 
in $\Bz\to\Kp\pim$ decays.

The time-dependent $\CP$-violating asymmetry in the decay $\Bz\to\pip\pim$ 
arises from interference between mixing and decay amplitudes, and interference 
between the tree and penguin decay amplitudes. A $\Bz\Bzb$ pair produced in 
\FourS\ decay evolves in time in a coherent $P$-wave state until one of the two 
mesons decays.  We reconstruct a sample of $B$ mesons ($\Bhh$) decaying to
the $\hh$ final state, where $h$ is a pion or kaon, and examine the remaining 
charged particles in each event to ``tag'' the flavor of the other $B$ meson (\Btag).  
The decay rate distribution $f_+\,(f_-)$ when $\hh = \pip\pim$
and $\Btag = \Bz\,(\Bzb)$ is given by~\cite{nirandquinn}
\begin{eqnarray}
\fpm = \frac{e^{-\left|\deltat\right|/\tau}}{4\tau} [1
& \pm & \spipi\sin(\deltamd\deltat) \nonumber \\
& \mp & \cpipi\cos(\deltamd\deltat)],
\label{fplusminus}
\end{eqnarray}
where $\tau$ is the $\Bz$ lifetime, $\deltamd$ is the $\Bz\Bzb$ mixing frequency, 
and $\deltat = t_{hh} - \ttag$ is the time between the \Bhh\ and \Btag\ decays.
The \CP-violating parameters $\spipi$ and $\cpipi$ are defined as
\begin{equation}
\spipi = \frac{2\,\ilam}{1+\alam^2}\quad{\rm and}\quad \cpipi = \frac{1-\alam^2}{1+\alam^2}.
\label{SandCdef}
\end{equation}
If the decay proceeds purely through the tree process $b\to uW^{-}$, the 
complex parameter $\lambda$ is directly related to CKM matrix elements,
\begin{equation}
\lambda(B\to\pip\pim) 
= \left(\frac{V_{tb}^*V_{td}}{V_{tb}V_{td}^*}\right)
\left(\frac{V_{ud}^*V_{ub}}{V_{ud}V_{ub}^*}\right),
\end{equation}
where we are assuming equal widths ($\Delta\Gamma_B = 0$) for the heavy and light
mass eigenstates.  Thus, at tree level in the Standard Model, $\alam = 1$ and 
$\ilam = \stwoa$, where $\alpha \equiv \arg\left[-V_{td}V_{tb}^*/V_{ud}V_{ub}^*\right]$.  

Recent theoretical estimates indicate that the contribution from the gluonic
penguin amplitude can be significant~\cite{Beneke01a,PQCD,Charming}.  The process 
$b\to dg$ carries the weak phase $\arg(V_{td}^*V_{tb})$, which can modify both 
the magnitude and phase of $\lambda$.  Thus, in general, $\alam\ne 1$ and 
$\ilam = \alam\sin{2\alpha_{\rm eff}}$,
where $\alpha_{\rm eff}$ depends on the magnitudes and strong phases of the tree and
penguin amplitudes.  Several approaches have been proposed to obtain 
information on $\alpha$ in the presence of penguins~\cite{Beneke01a,alphafrompenguins}.

In this analysis, we extract signal and background yields for $\pip\pim$, 
$\Kp\pim$, and $\Kp\Km$ decays~\cite{cc}, and the amplitudes of the $\pi\pi$ sine 
($\spipi$) and cosine ($\cpipi$) oscillation terms simultaneously from an unbinned maximum 
likelihood fit.  We parameterize the $K\pi$ component in terms of the total yield and 
the $\CP$-violating charge asymmetry
\begin{equation}
{\cal A}_{K\pi} \equiv \frac{N_{\Km\pip} - N_{\Kp\pim}}{N_{\Km\pip} + N_{\Kp\pim}}.
\label{kpiasym}
\end{equation}

The data sample used in this analysis consists of $33.7\invfb$
collected with the \babar\ detector at the SLAC PEP-II storage ring between 
October 1999 and June 2001.  The PEP-II facility operates nominally at the 
\Y4S\ resonance, providing collisions of $9.0\gev$ electrons on 
$3.1\gev$ positrons.  The data set includes $30.4\invfb$ collected in this 
configuration (on-resonance) and $3.3\invfb$ collected below the \BB\ threshold 
(off-resonance) that are used for continuum background studies.   

A detailed description of the \babar\ detector is presented in Ref.~\cite{ref:babar}.  
Charged particle (track) momenta are measured in a tracking system consisting 
of a 5-layer double-sided silicon vertex tracker (SVT) and a 40-layer drift 
chamber (DCH) filled with a gas mixture of helium and isobutane.  The SVT
and DCH operate within a $1.5\,{\rm T}$ superconducting solenoidal magnet.  
The typical decay vertex resolution for fully reconstructed $B$ decays is 
approximately $65\mum$ 
along the center-of-mass (CM) boost direction.  Photons are detected in an 
electromagnetic calorimeter (EMC) consisting of 6580 CsI(Tl) crystals arranged 
in barrel and forward endcap subdetectors.  The flux return for the solenoid
is composed of multiple layers of iron and resistive plate chambers for the 
identification of muons and long-lived neutral hadrons.

Tracks from the \Bhh\ decay are identified as pions or kaons by the 
Cherenkov angle $\theta_c$ measured with a detector of internally reflected 
Cherenkov light (DIRC).  The typical separation between pions and kaons varies 
from $8\sigma$ at $2\gevc$ to $2.5\sigma$ at $4\gevc$, where $\sigma$ is the 
average resolution on $\theta_c$.  Lower momentum kaons used in $B$ flavor 
tagging are identified with a combination of $\theta_c$ (for momenta down to 
$0.7\gevc$) and measurements of ionization energy loss $dE/dx$ in the DCH 
and SVT.

Hadronic events are selected based on track multiplicity and event
topology.  We require at least three tracks in the laboratory polar angle 
region $0.41 < \theta_{\rm lab} < 2.54$ satisfying the following requirements:
transverse momentum greater than $100\mevc$, at least $12$ DCH hits, and originating
from the interaction point within $10\cm$ in $z$ and $1.5\cm$ in $r$--$\varphi$~\cite{cs}.
Residual two-prong events from the reaction $\epem\to l^+l^-\,(l = e,\mu,\tau)$ 
are suppressed by requiring the ratio
of Fox-Wolfram moments $H_2/H_0$~\cite{fox} to be less than $0.95$ and
the sphericity~\cite{spheric} of the event to be greater than $0.01$.

Candidate $\B_{hh}$ decays are reconstructed from pairs of oppositely-charged 
tracks forming a good quality vertex, where the $\Bhh$ four-vector is calculated
assuming the pion mass for both tracks.  We require each track 
to have an associated $\theta_c$ measurement with a minimum of six Cherenkov 
photons above background, where the average is approximately 30 for both pions and 
kaons.  Protons are rejected based on $\theta_c$ and electrons 
are rejected based on $dE/dx$, shower shape in the EMC, and the ratio of shower 
energy and track momentum.  Background from the reaction 
$\epem\to q\bar{q}\; (q=u,d,s,c)$ is suppressed by removing jet-like events 
from the sample: we define the CM angle $\theta_S$ between the sphericity 
axes of the $B$ candidate and the remaining tracks and photons in the event, and require 
$\left|\cos{\theta_S}\right|<0.8$, which removes $83\%$ of the background.  
The total efficiency on signal events for all of the above selection is 
approximately $38\%$.

We define a beam-energy substituted mass 
$\mes = \sqrt{E^2_{\rm b}- {\mathbf {p}}_B^2}$.  The candidate energy is defined
as $E_{\rm b} =(s/2 + {\mathbf {p}}_i\cdot {\mathbf {p}}_B)/E_i$, 
where $\sqrt{s}$ and $E_i$ are the total energies of the \epem\ system in the
CM and laboratory frames, respectively, and ${\mathbf {p}}_i$ and ${\mathbf {p}}_B$ 
are the momentum vectors in the laboratory frame of the \epem\ system and the $\Bhh$
candidate, respectively.  Signal events are Gaussian distributed in $\mes$ 
with a mean near the $B$ mass and a resolution of $2.6\mevcc$, dominated by
the beam energy spread.  The background shape is parameterized by a threshold 
function~\cite{ARGUS} with a fixed endpoint given by the average beam energy.

We define a second kinematic variable $\de$ as the difference between the energy
of the $\Bhh$ candidate in the CM frame and $\sqrt{s}/2$.  The $\de$ distribution is 
peaked near zero for $\pip\pim$ decays.  For decays with one\,(two) kaons, the
distribution is shifted relative to $\pi\pi$ on average by $-45\mev$ ($-91\mev$), 
respectively, where the exact separation depends on the laboratory momentum of
the kaon(s).  The resolution on $\de$ for signal decays is approximately $26\mev$.  
The background is parameterized by a quadratic function.  

Candidate $\hh$ pairs selected in the region $5.2 < \mes < 5.3\gevcc$ 
and $\left|\de\right|<0.15\gev$ are used to extract yields and \CP-violating 
asymmetries with an unbinned maximum likelihood fit.  The total number of events in 
the fit region satisfying all of the above criteria is $9741$.  A sideband region, 
defined as $5.20 < \mes < 5.26\gevcc$ and $\left|\de\right|<0.42\gev$, is used 
to extract various background parameters.

The analysis method combines the techniques used to measure charmless two-body
branching fractions~\cite{twobodyPRL} and $\stwob$~\cite{BabarSin2betaObs}.  
The primary issues in this analysis are determination of the $\Btag$ flavor, 
measurement of the distance $\deltaz$ between the \Bhh\ and \Btag\ decay vertices, 
discrimination of signal from background, identification of pions and kaons, and 
extraction of yields and \CP\ asymmetries.

To determine the flavor of the \Btag\ meson we use the same $B$-tagging algorithm used 
in the $\stwob$ and $\Bz$--$\Bzb$ mixing~\cite{ref:babarmix} analyses.  The algorithm 
relies on the correlation between the flavor of the $b$ quark and the charge of the 
remaining tracks in the event after removal of the \Bhh\ candidate.  We define five 
mutually exclusive tagging categories: {\tt Lepton}, {\tt Kaon}, {\tt NT1}, {\tt NT2}, and 
{\tt Untagged}.  {\tt Lepton} tags rely on primary electrons and muons from semileptonic 
$B$ decays, while {\tt Kaon} tags exploit the correlation in the process $b\to c\to s$ 
between the net kaon charge and the charge of the $b$ quark.  
The {\tt NT1} and {\tt NT2} categories are derived from a neural network that is sensitive 
to charge correlations between the parent \B\ and unidentified leptons and kaons, soft pions, 
or the charge and momentum of the track with the highest CM momentum.  The addition of 
{\tt Untagged} events improves the signal yield estimates and provides a larger 
sample for determining background shape parameters directly in the maximum likelihood fit.

The quality of tagging is expressed in terms of the effective efficiency 
$Q = \sum_i \epsilon_i D_i^2$, where $\epsilon_i$ is the fraction of events tagged in 
category $i$ and the dilution $D_i = 1-2w_i$ is related to the mistag fraction $w_i$.  
The statistical errors on $\spipi$ and $\cpipi$ are proportional to $1/\sqrt{Q}$.  Table~\ref{tab:tagging} 
summarizes the tagging performance in a data sample \Bflav\ of fully reconstructed neutral $B$ decays 
into $D^{(*)-}h^+\,(h^+ = \pip, \rho^+, a_1^+)$ and
$\jpsi K^{*0}\,(K^{*0}\to\Kp\pim)$ flavor eigenstates
We use the same tagging efficiencies and dilutions for signal $\pi\pi$, $K\pi$, 
and $KK$ decays.  Separate background tagging efficiencies for each species are 
obtained from a fit to the $\hh$ on-resonance sideband data and reported in 
Table~\ref{tab:bkgtag}.  

\begin{table}[!tbp]
\caption{Tagging efficiency $\epsilon$, average dilution $D = 1/2\left(D_{\Bz} + D_{\Bzb}\right)$, 
dilution difference $\diffD = D_{\Bz} - D_{\Bzb}$, and effective tagging efficiency $Q$
for signal events in each tagging category.}
\smallskip
\begin{center}
\begin{tabular}{ccccc} \hline\hline
Category & $\epsilon\,(\%)$ & $D\,(\%)$ & $\diffD\,(\%)$ & $Q\,(\%)$ \rule[-2mm]{0mm}{6mm} \\\hline
{\tt Lepton}   & $11.0\pm 0.3$ & $82.3 \pm 2.7$ & $-2.1  \pm 4.5$ & $7.5\pm  0.5$ \rule[-1.5mm]{0mm}{5mm}\\
{\tt Kaon}     & $35.8\pm 0.5$ & $64.8 \pm 2.0$ & $ 3.5  \pm 3.1$ & $15.0\pm 1.0$ \rule[-1.5mm]{0mm}{4mm}\\
{\tt NT1}      & $8.0 \pm 0.3$ & $55.6 \pm 4.2$ & $-12.1 \pm 6.7$ & $2.5\pm  0.4$ \rule[-1.5mm]{0mm}{1.5mm}\\
{\tt NT2}      & $13.9\pm 0.4$ & $30.2 \pm 3.8$ & $ 9.0  \pm 5.7$ & $1.3\pm  0.3$ \rule[-1.5mm]{0mm}{1.5mm}\\
{\tt Untagged} & $31.3\pm 0.5$ & -- 	  & --            & --          \rule[-1.5mm]{0mm}{1.5mm}\\ \hline
Total $Q$ & & & & $26.3\pm 1.2$ \rule[-2mm]{0mm}{6mm} \\\hline\hline
\end{tabular}
\end{center}
\label{tab:tagging}
\end{table}

\begin{table}[!tbp]
\caption{Tagging efficiencies $(\%)$ for background events in each species.}
\smallskip
\begin{center}
\begin{tabular}{cccccc} \hline\hline
Category & $\epsilon(\pi\pi)$ & $\epsilon(K\pi)$ & $\epsilon(KK)$ \rule[-2mm]{0mm}{6mm} \\\hline 
{\tt Lepton}   & $1.0\pm 0.1$  & $1.0\pm 0.1$  & $1.5\pm 0.2$  \rule[-1.5mm]{0mm}{5mm}\\
{\tt Kaon}     & $26.0\pm 0.4$ & $33.1\pm 0.6$ & $23.5\pm 0.7$ \rule[-1.5mm]{0mm}{1.5mm}\\
{\tt NT1}      & $6.6\pm 0.2$  & $5.4\pm 0.3$  & $6.9\pm 0.4$  \rule[-1.5mm]{0mm}{1.5mm}\\
{\tt NT2}      & $17.6\pm 0.4$ & $15.3\pm 0.5$ & $19.7\pm 0.6$ \rule[-1.5mm]{0mm}{1.5mm}\\
{\tt Untagged} & $48.9\pm 0.7$ & $45.2\pm 0.6$ & $48.3\pm 0.8$\rule[-2mm]{0mm}{3mm} \\\hline\hline
\end{tabular}
\end{center}
\label{tab:bkgtag}
\end{table}

The time difference $\deltat$ is obtained from the measured distance between 
the $z$ position of the $\Bhh$ and $\Btag$ decay vertices and the known boost 
of the $\epem$ system.  The $z$ position of the \Btag\ vertex is determined 
with an iterative procedure that removes tracks with a large contribution to 
the total $\chi^2$~\cite{BabarSin2betaObs,ref:babarmix}.  An additional 
constraint is constructed from the three-momentum and vertex position of the 
\Bhh\ candidate, and the average $\epem$ interaction point and boost.  The 
typical $\deltaz$ resolution is $180\mum$.  We require 
$\left|\deltat\right|<17\ps$ and $0.3 < \sigma_{\deltat} < 3.0\ps$, where 
$\sigma_{\deltat}$ is the error from the vertex fit.  The resolution function 
for signal candidates is a sum of three Gaussians, identical to the one 
described in Ref.~\cite{BabarSin2betaObs}, with parameters determined from a 
fit to the \Bflav\ sample (including events in all five tagging categories).  
The background resolution function is parameterized as the sum of three 
Gaussians, with the parameters determined from a fit to the $\hh$ on-resonance 
sideband data.  

The data sample used in the fit contains $97\%$ background, mostly due to random
combinations of tracks produced in $\epem\to q\bar{q}$ events.
Discrimination of signal from background in the maximum likelihood fit is enhanced
by the use of a Fisher discriminant ${\cal F}$~\cite{twobodyPRL}.  The discriminating
variables are constructed from the scalar sum of the CM momenta of all tracks and photons 
(excluding tracks from the \Bhh\ candidate) entering nine two-sided $10$-degree concentric 
cones centered on the thrust axis of the \Bhh\ candidate.  
The distribution of ${\cal F}$ for signal events is parameterized as a single Gaussian, 
with parameters determined from Monte Carlo simulated decays and validated with 
$\Bub\to\Dz\pim$ decays reconstructed in data.  The background shape is parameterized as 
the sum of two Gaussians, with parameters determined directly in the maximum likelihood fit.

Identification of $\hh$ tracks as pions or kaons is accomplished with the 
Cherenkov angle measurement from the DIRC.  We construct Gaussian probability density functions 
(PDFs) from the difference between measured and expected values of $\theta_c$ for the pion or 
kaon hypothesis, normalized by the resolution.  The DIRC performance is parameterized using a 
sample of $D^{*+}\to\Dz\pip$, $\Dz\to \Km\pip$ decays reconstructed in data.  
Within the statistical precision of the control sample (approximately $10^5$ events), we find 
similar response for positively and negatively charged tracks and use a single parameterization 
for both.  

We use an unbinned extended maximum likelihood fit to extract yields and $\CP$ parameters
from the $\Bhh$ sample.  The likelihood for candidate $j$ tagged in category 
$c$ is obtained by summing the product of event yield $n_{i}$, tagging efficiency $\epsilon_{i,c}$,
and probability ${\cal P}_{i,c}$ over the eight possible signal and background hypotheses $i$
(referring to $\pi\pi$, $\Kp\pim$, $\Km\pip$, and $KK$ decays),

\begin{equation}
{\cal L}_c = \exp{\left(-\sum_{i}n_i\epsilon_{i,c}\right)}
\prod_{j}\left[\sum_{i}n_i\epsilon_{i,c}{\cal P}_{i,c}(\vec{x}_j;\vec{\alpha}_i)\right].
\end{equation}
For the $K^{\mp}\pi^{\pm}$ hypotheses, the yield is parameterized as 
$n_i = N_{K\pi}\left(1 \pm {\cal A}_{K\pi}\right)/2$, where 
$N_{K\pi} = N_{\Km\pip} + N_{\Kp\pim}$.
We fix the tagging efficiencies $\epsilon_i$ to the values in Tables~\ref{tab:tagging} and
\ref{tab:bkgtag}.  The probabilities ${\cal P}_{i,c}$ are evaluated as the product of PDFs 
for each of the independent variables 
$\vec{x}_j = \left\{\mes, \de, {\cal F}, \theta_c^+, \theta_c^-, \deltat\right\}$, where $\theta_c^+$
and $\theta_c^-$ are the Cherenkov angles for the positively and negatively charged tracks.  The 
total likelihood ${\cal L}$ is the product of likelihoods for each tagging category and the 
free parameters are determined by minimizing the quantity $\ln{\cal L}$.

The $\deltat$ PDF for signal $\pip\pim$ decays is given by Eq.~\ref{fplusminus}, modified to
include the dilution and dilution difference for each tagging category, and convolved with 
the signal resolution function.  The $\deltat$ PDF for signal $K\pi$ events takes into account
$\Bz$--$\Bzb$ mixing,
depending on the charge of the kaon and the flavor of $\Btag$.  We parameterize
$\Bz\to\Kp\Km$ decays as an exponential convolved with the resolution function.

There are $18$ free parameters in the fit.  In addition to the \CP-violating parameters
$\spipi$, $\cpipi$, and ${\cal A}_{K\pi}$, the fit determines signal and background yields
(six parameters), the background $K\pi$ charge asymmetry, and eight parameters describing
the background shapes in $\mes$, $\de$, and ${\cal F}$.  We fix $\tau$ and $\deltamd$ to 
the world-average values~\cite{PDG2000}.

In a sample of $33$ million $\BB$ pairs we find $65^{+12}_{-11}$ $\pi\pi$, $217\pm 18$ $K\pi$, and
$4.3^{+6.3}_{-4.3}$ $KK$ events.  These yields are consistent with the branching fractions reported in
Ref.~\cite{twobodyPRL}, as well as measurements from other experiments~\cite{CLEOTwoBody,BELLETwoBody}.  
The results for \CP-violating asymmetries are 
summarized in Table~\ref{tab:results}.  Statistical errors correspond to unit change in 
$\chi^2 \equiv -2\ln({\cal L})$.  For each parameter, we also calculate the $90\%$ confidence 
level (C.L.) interval corresponding to a change in $\chi^2$ of $2.69$, and taking into account
the systematic error.  The correlation between $\spipi$ and $\cpipi$ is $-21\%$, while 
${\cal A}_{K\pi}$ is uncorrelated with either $\spipi$ or $\cpipi$.

\begin{table}[!tbp]
\caption{Central values and $90\%$ C.L. intervals for $\spipi$, $\cpipi$,
and ${\cal A}_{K\pi}$ from the maximum likelihood fit.}
\smallskip
\begin{center}
\begin{tabular}{ccc} \hline\hline
Parameter & Central Value & $90\%$ C.L. Interval \rule[-2mm]{0mm}{6mm} \\\hline  \rule[-2mm]{0mm}{6mm} 
$\spipi$            & $0.03^{+0.53}_{-0.56}\pm 0.11$  & $\left[-0.89,+0.85\right]$ \\
$\cpipi$            & $-0.25^{+0.45}_{-0.47}\pm 0.14$ & $\left[-1.0,+0.47\right]$  \\
${\cal A}_{K\pi}$   & $-0.07 \pm 0.08 \pm 0.02$       & $\left[-0.21,+0.07\right]$ \rule[-2mm]{0mm}{5mm} \\\hline\hline
\end{tabular}
\end{center}
\label{tab:results}
\end{table}

\begin{figure}[!tbp]
\begin{center}
\begin{minipage}[h]{4.75cm}
\includegraphics[width=4.75cm]{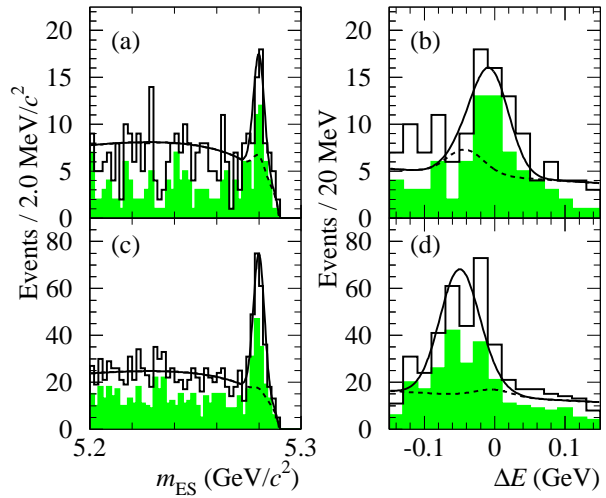}
\end{minipage}
\end{center}
\caption{Distributions of $\mes$ and $\de$ (unshaded histograms)
for events enhanced in signal (a), (b) $\pi\pi$ and (c), (d) $K\pi$ 
decays based on the 
likelihood ratio selection described in the text.  Solid curves represent 
projections of the maximum likelihood fit result after accounting for the 
efficiency of the additional selection, while dashed curves represent 
$q\bar{q}$ and $\pi\pi\leftrightarrow K\pi$ cross-feed background.  Shaded 
histograms show the subset of events that are tagged.}
\label{fig:prplots}
\end{figure}

Figure~\ref{fig:prplots} shows distributions of $\mes$ and 
$\de$ for events enhanced in signal decays based on likelihood ratios.  We define
${\cal R}_{\rm sig} = \sum_s{n_s{\cal P}_s}/\sum_i{n_i{\cal P}_i}$ and 
${\cal R}_k = n_k{\cal P}_k/\sum_s{n_s{\cal P}_s}$, where $\sum_s\, (\sum_i)$ indicates a sum 
over signal\,(all) hypotheses, and ${\cal P}_k$ indicates the probability for signal
hypothesis $k$.  The probabilities include the PDFs for $\theta_c$, ${\cal F}$, and 
$\mes\,(\Delta E)$ when plotting $\Delta E\,(\mes)$.  The selection is defined
by optimizing the signal significance with respect to ${\cal R}_{\rm sig}$ and ${\cal R}_k$.
The solid curve in each plot represents the fit projection after correcting for the
efficiency of the additional selection (approximately $55\%$ for $\pi\pi$ and 
$85\%$ for $K\pi$).  

\begin{figure}[!tbp]
\begin{center}
\begin{minipage}[h]{3.75cm}
\includegraphics[width=3.75cm]{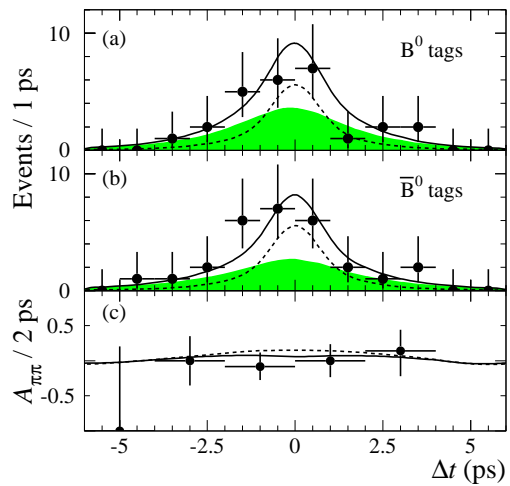}
\end{minipage}
\end{center}
\caption{Distributions of $\deltat$ for events enhanced in 
signal $\pi\pi$ decays based on the likelihood ratio selection described in the 
text.  Figures (a) and (b) show events (points with errors) with $\Btag=\Bz$ or 
$\Bzb$.  Solid curves represent projections of the maximum 
likelihood fit, dashed curves represent the sum of $q\bar{q}$ and $K\pi$ 
background events, and the shaded region represents the contribution from
signal $\pi\pi$ events.  Figure (c) shows ${\cal A}_{\pi\pi}(\deltat)$ for data 
(points with errors), as well as fit projections for signal and background 
events (solid curve), and signal events only (dashed curve).}
\label{fig:dtplot}
\end{figure}

Figure~\ref{fig:dtplot} shows the $\deltat$ distributions and the asymmetry
${\cal A}_{\pi\pi}(\deltat) = 
(N_{\Bz}(\deltat) - N_{\Bzb}(\deltat))/(N_{\Bz}(\deltat) + N_{\Bzb}(\deltat))$
for tagged events enhanced in signal $\pi\pi$ decays.  The selection procedure is the
same as Fig.~\ref{fig:prplots}, with the likelihoods defined including the PDFs for 
$\theta_c$, ${\cal F}$, $\mes$, and $\de$.  Approximately $24$ $\pi\pi$, $22$ $q\bar{q}$, 
and $5$ $K\pi$ events satisfy the selection.

Systematic uncertainties on $\spipi$, $\cpipi$, and ${\cal A}_{K\pi}$ arise primarily
from imperfect knowledge of the PDF shapes and uncertainties on tagging efficiencies, 
dilutions, $\tau$, and $\deltamd$.  The total systematic error is calculated 
as the sum in quadrature of the individual uncertainties.  The error on ${\cal A}_{K\pi}$ is
dominated by uncertainty in the mean of the $\de$ PDF ($0.01$) and possible charge bias
in track and $\theta_c$ reconstruction ($0.01$)~\cite{babardircp}.  Errors on $\spipi$ and $\cpipi$
are dominated by the parameterization of $\deltat$ resolution for signal and background
($\approx 0.07$ for $\spipi$, $\approx 0.03$ for $\cpipi$), tagging ($0.05$), and, for
$\cpipi$ only, the mean of the $\de$ PDF ($0.1$).

Extensive studies were performed to validate the fit technique.  A large ensemble of 
Monte Carlo pseudo-experiments was generated from the nominal PDFs with the statistics 
observed in the full data set.  Parameter errors and the maximum value of the likelihood
obtained in the data fit are all consistent with expectations based on these 
pseudo-experiments, and all free parameters are unbiased.  We have checked that 
consistent results are obtained when separating events by $\Btag$ flavor.  As a validation of 
the $\deltat$ parameterization in data, we fit the full data set to simultaneously extract 
yields, background parameters, $\tau$, $\deltamd$, $\spipi$, and $\cpipi$.  We find 
$\tau = (1.52\pm 0.12)\ps$ and $\deltamd = (0.54\pm 0.09)\hbar \ps^{-1}$, and all
other parameters are consistent with the nominal fit.

In summary, we have presented a measurement of time-dependent $\CP$-violating
asymmetries in $\Bz\to\pip\pim$ decays and an updated measurement of 
the charge asymmetry ${\cal A}_{K\pi}$.  The latter is consistent with our previous 
result reported in Ref.~\cite{twobodyPRL}, as well as results from other
experiments~\cite{CLEOdircp,BELLEdircp}.  
We observe no evidence for direct \CP\ violation in the $K\pi$ mode and determine a 
$90\%$ C.L. interval excluding a significant part of the allowed region.  
Although the current measurements of $\spipi$ and 
$\cpipi$ do not significantly constrain the Unitarity Triangle, with the addition of more 
data and further improvements in detector performance and analysis techniques, future 
results will yield important information about \CP\ violation in the $B$-meson system.

We are grateful for the excellent luminosity and machine conditions
provided by our \pep2\ colleagues.
The collaborating institutions wish to thank 
SLAC for its support and kind hospitality. 
This work is supported by
DOE
and NSF (USA),
NSERC (Canada),
IHEP (China),
CEA and
CNRS-IN2P3
(France),
BMBF
(Germany),
INFN (Italy),
NFR (Norway),
MIST (Russia), and
PPARC (United Kingdom). 
Individuals have received support from the Swiss NSF, 
A.~P.~Sloan Foundation, 
Research Corporation,
and Alexander von Humboldt Foundation.


\begin{thebibliography}{99}
\bibitem{cc}
When only one charge mode is given, conjugate decay modes are implied.
\bibitem{cs}
The unit vector $\hat{z}$ is aligned along the detector axis in
the electron beam direction.
\end{thebibliography}
\end{document}